\pdfoutput=1

\documentclass[useAMS,usenatbib]{mn2e}

\usepackage{pifont}
\usepackage[final]{graphicx}
\usepackage{amssymb}
\usepackage{color}
\usepackage{url}

\title[Radiation feedback on dusty clouds during Seyfert activity]{Radiation feedback on dusty clouds during Seyfert activity}
\author[M. Schartmann et al.]
  {M.~Schartmann$^{1,2,}$\thanks{E-mail: schartmann@mpe.mpg.de},
   M.~Krause$^{1,2}$,
   and A.~Burkert$^{1,2,}$\thanks{Max Planck Fellow},\\
$^{1}$Max-Planck-Institut f\"ur extraterrestrische Physik, Giessenbachstra\ss e, D-85748 Garching, Germany\\
$^{2}$Universit\"ats-Sternwarte M\"unchen, Scheinerstra\ss e 1, D-81679 M\"unchen, Germany}
\begin{document}

\date{Accepted . Received ; in original form }

\pagerange{\pageref{firstpage}--\pageref{lastpage}} \pubyear{2002}

\maketitle

\label{firstpage}

\begin{abstract}
We investigate the evolution of dusty gas clouds falling into the centre of an active
Seyfert nucleus. Two-dimensional 
high-resolution radiation hydrodynamics simulations are performed to study the fate
of single clouds and the interaction between two clouds approaching the Active Galactic Nucleus.
We find three distinct phases of the evolution of the cloud: (i) formation of a lenticular 
shape with dense inner rim caused by the interaction of gravity and radiation pressure 
(the {\it lense phase}), (ii) formation of a clumpy sickle-shaped structure as 
the result of a converging flow (the {\it clumpy sickle phase}) and 
(iii) a {\it filamentary phase} caused by
a rapidly varying optical depth along the sickle. 
Depending on the column density of the cloud, 
it will either be pushed outwards or its central (highest column density) 
parts move inwards,
while there is always some material pushed outwards by radiation pressure effects. 
The general dynamical evolution of the cloud 
can approximately be described by a simple analytical model.
\end{abstract}


\section{Introduction}
\label{sec:introduction}

Normal spiral galaxies light up as soon as enough gas is accreted onto their nuclei. 
Then, their central region becomes similarly bright as the stars of the whole galaxy, a phenomenon
called {\it Seyfert activity} \citep{Seyfert_43,Weedman_77,Sanders_81}. It is thought that this is a recurrent process 
and most of the normal galaxies have encountered such activity cycles during the growth phase of their central
supermassive black holes, whenever enough gas 
reaches their centres. A fast rotating, thin and hot gaseous accretion disc forms,  
which is surrounded by a ring-like, dusty, geometrically thick gas 
reservoir -- the so-called {\it dusty torus}. 
Anisotropically blocking the light, this gives rise to two characteristic observational signatures, depending 
whether the line of sight is obscured (edge-on view, so-called {\it Seyfert~2 galaxies}) or not (face-on view, 
so-called {\it Seyfert~1 galaxies}). 
These nuclear regions of nearby Seyfert 
galaxies, as well as our own galactic centre have been observed in great detail with the most up-to-date instruments at 
the largest available telescopes and interferometers, yielding unprecedented resolution
\citep[e.~g.~][]{Davies_07,Tristram_07,Tristram_09,Burtscher_09,Prieto_10,Hoenig_10}.
Therefore, they represent an ideal testbed for studying fueling processes and
the characteristics of active galactic nuclei (AGN).
Only a few models capable of explaining the necessary fueling process have been 
presented up to today.
For example \citet{Elitzur_06} argue for fueling of the central region through a midplane influx of 
cold and clumpy material from the galaxy \citep{Shlosman_90}. When reaching the centre, the gas will 
contribute to the formation of a hot accretion disc that illuminates the surrounding region. 
A hydromagnetically or radiatively driven disc wind forms. The embedded dusty and optically 
thick clouds then form the {\it Toroidal Obscuration Region (TOR)}
in the parsec scale vicinity of the central engine, which replaces the classical
torus in their model. 
The accretion process from galactic scales down to sub-parsec scales has also been followed 
in great detail by \citet{Hopkins_10} in multiscale SPH (smoothed particle hydrodynamics) 
simulations taking gas, stars, black holes, star formation and stellar feedback into account. 
Accretion rates up to a few solar masses per year can be obtained. 
Detailed simulations of gas clump interactions with the central super-massive black hole (SMBH)
in the Galactic Centre
have been performed for example by \citet{Bonnell_08}, \citet{Hobbs_09} and \citet{Alig_11}.
They find that this process can lead to the formation of a compact gaseous accretion disc, which
might be the progenitor of one of the stellar discs observed. By efficiently redistributing
angular momentum when such a cloud overlaps the black hole, \citet{Alig_11} find that 
this process might as well result in a period of Seyfert activity. \\

\begin{figure}
\begin{center}
\includegraphics[width=0.65\linewidth]{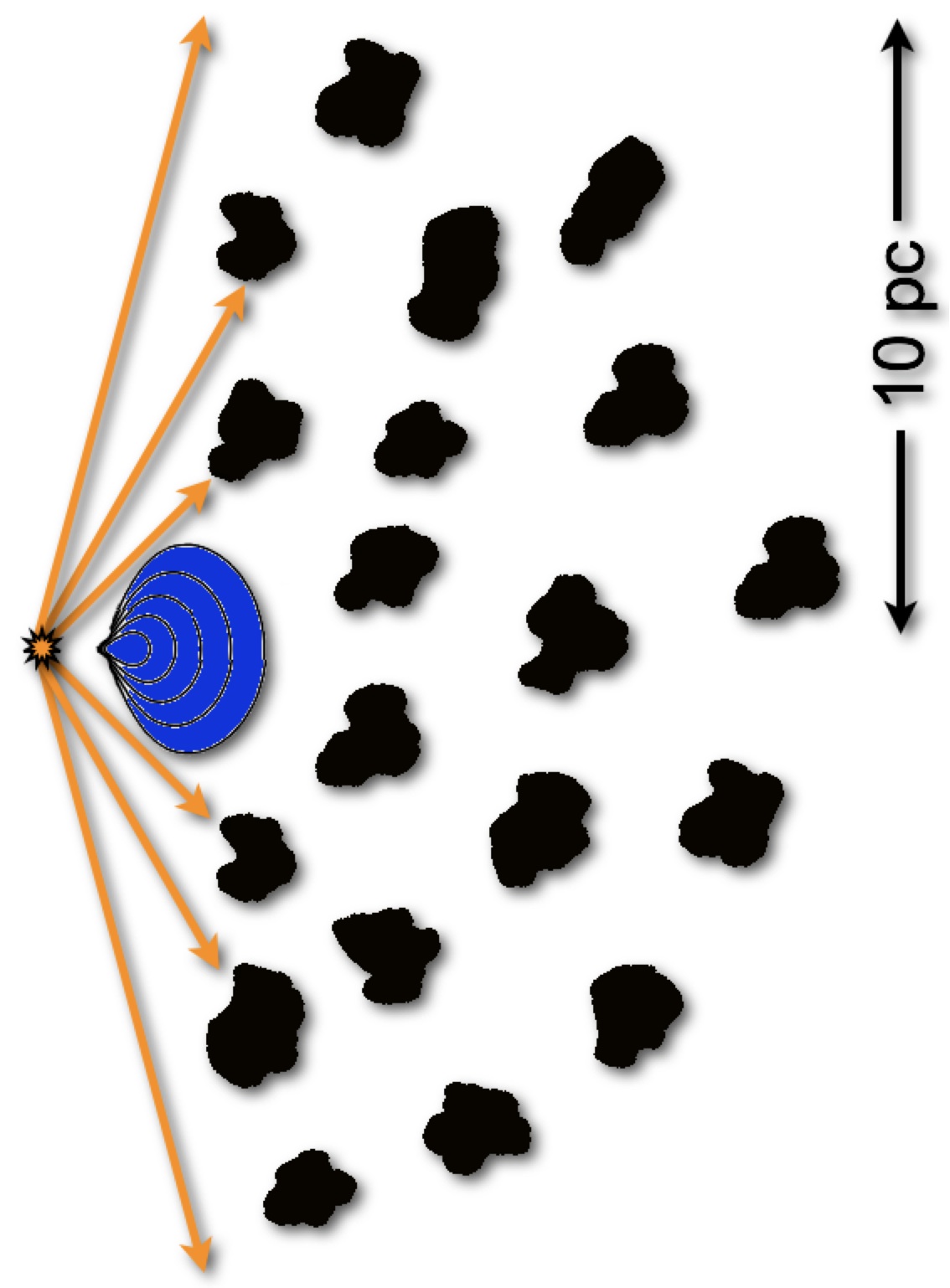}
\caption{Sketch of the central region of a Seyfert nucleus as found in the 3D simulations done in \citet{Schartmann_10}.}
\label{fig:clumpy_torus_sketch}
\end{center}
\end{figure}

In this article, we will concentrate on a model which links the evolution of young and massive nuclear star clusters
to the evolution of the central engine. The implications of stellar mass-loss 
within the nuclear star clusters of quasars 
have been investigated with the help of analytical considerations for example by \citet{Norman_88} 
and \citet{Scoville_88,Scoville_95}. 
They find that stellar processes play an important role for the fuelling of the central black hole and
the envelopes of giant stars might as well correspond to the clouds in the Broad Line Regions ({\it BLR})
of galactic nuclei. Observations of nearby Seyfert galaxies indeed find evidence for young and massive
nuclear star clusters and a tentative connection with the onset of nuclear activity \citep{Davies_07}.
\citet{Schartmann_09,Schartmann_10} are able to confirm this idea with the help 
of detailed hydrodynamical simulations. 
During the {\it Asymptotic Giant Branch} (AGB) phase of the evolution of the nuclear star cluster, slow stellar winds provide enough 
low angular momentum fuel, which can be accreted towards the central region to explain their 
observed core luminosities, amongst other observational properties \citep{Schartmann_10}.
The typical outcome of such a simulation is a two-component structure: 
(i) a filamentary or clumpy stream of gas, which feeds clumps towards the
centre from the tens of parsec scale vicinity of the black hole and (ii) a geometrically thin 
accretion disc around the SMBH on sub-parsec to parsec 
scale \citep{Schartmann_09}. With the help of a one-dimensional 
effective treatment of the central few parsecs including the effects of rotation, viscosity, mass inflow from large
scales and star formation, \citet{Schartmann_10} are able to show that a significant amount of matter in the disc can be 
accreted towards the centre. Finally reaching the vicinity of the black hole, this will lead to the formation of 
a hot inner accretion disc and
the birth of the AGN. The outer parsec-sized disc of gas and dust (potentially already puffed up to a toroidal shape 
by a thus far unknown 
physical process) will shield part of the radiation. A transition between a completely 
shadowed region (behind the torus midplane)
and a region exerted to full radiation pressure from the source (around the polar axis) is expected 
(as sketched in Fig.~\ref{fig:clumpy_torus_sketch}). 
The 3D models in \citet{Schartmann_09,Schartmann_10}
which solely cover the pre-active phase, where radiative pressure forces from the 
central source are negligible, produce clouds in both of these two regions. 
To investigate the feedback properties transmitted by the radiation pressure of the hot accretion disc in the active 
phase and how it affects infalling clouds is the subject of this work (see Fig.~\ref{fig:clumpy_torus_sketch}). 

In Sect.~\ref{sec:mod_num}, we describe the numerical model and physical setup of our simulations and explain the 
test problems we used to assess their accuracy. Sect.~\ref{sec:results} describes our simulations and the main
results of our parameter studies. It is followed by a critical discussion in Sect.~\ref{sec:discussion},  
before we conclude in Sect.~\ref{sec:conclusions}.

\section{Model and test calculations}
\label{sec:mod_num}

\subsection{The approximative radiative transfer approach}
\label{sec:radtrans}

Even with today's computational power the simultaneous solution of the 
hydrodynamical evolution and the time-dependent 
radiative transfer equation is impractical for high-resolution multi-dimensional simulations.
Severe simplifications have to be applied, depending on the problem under investigation.

In the simulations shown here, we explore the situation of a strong point source, illuminating 
a spatially confined cloud, immerged at a distance of several tens of dust sublimation 
radii in initially dust-free, 
low density gas. 
The spectral energy distribution of the central source peaks in 
the ultra-violet wavelength regime, where also the mass absorption coefficient of the 
adopted dust model shows a prominent maximum. Therefore, radiation can only penetrate a relatively
short distance into the cloud, before the high energy UV-photons get absorbed and
re-emitted in the infrared wavelength regime. By consequence, the surface of the cloud in the direction 
of the central source will receive almost all of the radiative acceleration. 
Given the steep drop of the dust temperature distribution at this rim and the fact that 
the clouds are far away from the sublimation radius, 
secondary infrared radiation pressure effects are of minor importance
for the dynamical evolution in our infalling clump scenario.
A second consequence is that radiation pressure effects predominantly act in radial direction and
are dynamically unimportant in vertical direction, as long as the dust temperatures are low
and the optical depth is not too high. 
Hence, a one-dimensional treatment of the 
radiative transfer problem is reasonable here. 
Furthermore, we are mainly interested in the dynamical evolution and not in the detailed 
thermodynamics of the dust distribution. 

The issues raised above justify the following simplified approach:
We treat the central accretion disc as an isotropically radiating point-source
with a spectral energy distribution as shown in Fig.~3b of \citet{Schartmann_05}, but normalised to 
correspond to 10\% of the Eddington luminosity for the case of the nearby Seyfert\,2 
galaxy NGC\,1068. 
The radiation is divided into 54 wavelength bins and propagated along radial rays 
outwards, where we take geometrical dilution and absorption and reemission by the dust grains in 
each cell into account. Scattering is neglected. A full radiative transfer 
calculation within each time step is done. 
We further make a one fluid assumption and fully couple gas and dust dynamically. 
The reason for this coupling is that dust grains are charged due to the UV and X-ray radiation
and couple to the gas with the help of magnetic fields. The effects of grain charging and gas-dust-coupling
have been investigated e.~g.~by \citet{Scoville_95}.
We calculate the gas and dust temperatures separately by assuming that
no heat is transferred between the gas and the dust phase.
Only those cells receive accelerating forces, which possess gas temperatures below 
a threshold temperature $T_{\mathrm{sputt}}=10^5$K. At this temperature, the rates of change of the
grain radii of silicate and graphite grains show a steep rise \citep{Dwek_96}, caused 
by sputtering processes transmitted by the hot gas, they are embedded in. 
From $T_{\mathrm{sputt}}$ onwards, we assume that the hot gas will
destroy the dust content of the given cell instantaneously. Those cells with 
a density below a gas density threshold $\rho_{\mathrm{gas}}^{\mathrm{thresh}}$ will not 
be accelerated either (typically chosen to be twice the minimum density threshold 
of the simulation, see Table~\ref{tab:params}). Otherwise, this 
would lead to artificial generation of matter at the inner boundary of the domain
due to the lower limit of the gas density in the simulations. These
criteria enable us to distinguish between the gas and the dust phase.

\subsection{Opacity model}
\label{sec:opac_model}

We use a standard galactic dust model, which is split into 54 frequency bins 
and has averaged dust grain properties. 
Grain radii vary between 
0.005\,$\mu$m and 0.25\,$\mu$m with a number density distribution proportional to 
$a^{-3.5}$, where $a$ is the grain radius \citep{Mathis_77}. It comprises of
62.5\% of silicate and 37.5\% of graphite grains, where for the latter, the anisotropic behaviour
is taken into account. Optical constants are adopted from \citet{Draine_84}, \citet{Laor_93} and 
\citet{Weingartner_01}. The resulting opacity curve is shown as the dashed line in Fig.~3a of 
\citet{Schartmann_05}. A gas-to-dust mass ratio of 150 is used in the simulations 
shown in this paper.

\subsection{Test simulations}

In order to demonstrate the suitability of our numerical approach of simulating radiative effects, 
we performed a number of test simulations. The most relevant will be discussed in this section.

\subsubsection{Radiative transfer}
\label{sec:radtranstest}

To test the accuracy and validity of our simplified radiative transfer treatment, 
we compare the resulting temperature distributions with the three-dimensional 
Monte-Carlo radiative transfer code RADMC-3D \citep{Dullemond_10}. 
The computational domain is chosen to have the same physical size and resolution 
as the cloud simulations described in this paper, but 
we fill it homogeneously with dust with densities of 
$\rho_{\mathrm{d}}=10^{-25}\,\mathrm{g}\,\mathrm{cm}^{-3}$, 
$\rho_{\mathrm{d}}=10^{-24}\,\mathrm{g}\,\mathrm{cm}^{-3}$, 
$\rho_{\mathrm{d}}=10^{-23}\,\mathrm{g}\,\mathrm{cm}^{-3}$
and $\rho_{\mathrm{d}}=10^{-22}\,\mathrm{g}\,\mathrm{cm}^{-3}$.
These correspond to optical depths at a wavelength of 
$\lambda=9.8\,\mu$m of $\tau_{9.8\mu\mathrm{m}}=0.05$, $0.50$,
4.97 and 49.73, where a dust absorption coefficient 
of $\kappa_{\mathrm{abs}}^{9.8\mu\mathrm{m}}=16,446.66\,\mathrm{cm}^2\mathrm{g}^{-1}$ was used, 
taken from our frequency dependent dust model, as described in Sect.~\ref{sec:opac_model}. 

\begin{figure}
\begin{center}
\includegraphics[width=0.9\linewidth]{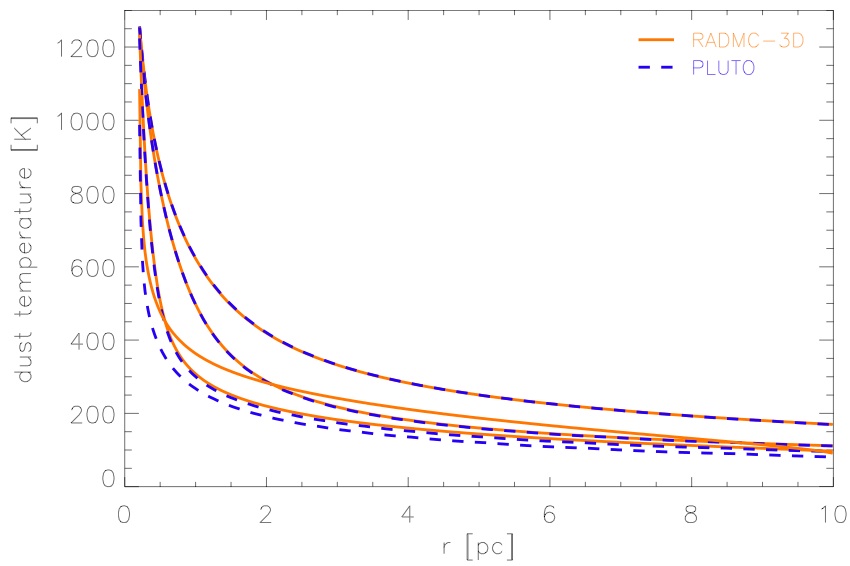}
\caption{Comparison of the radial dust temperature distribution for an optical depth varying between
         $\tau_{9.8\mu\mathrm{m}}=0.05$ (uppermost blue dashed curve) to 
         $\tau_{9.8\mu\mathrm{m}}=49.73$ (lowermost blue dashed curve) 
         in steps of a factor of 10 within a homogeneous, spherical distribution of dust.
         The blue dashed lines show the result of our approximate radiative transfer solver, whereas the orange solid lines 
         represent the reference solution calculated with RADMC-3D \citep{Dullemond_10}.}
\label{fig:plot_comp_radmc3d_pluto}
\end{center}
\end{figure}

\begin{figure}
\begin{center}
\includegraphics[width=0.9\linewidth]{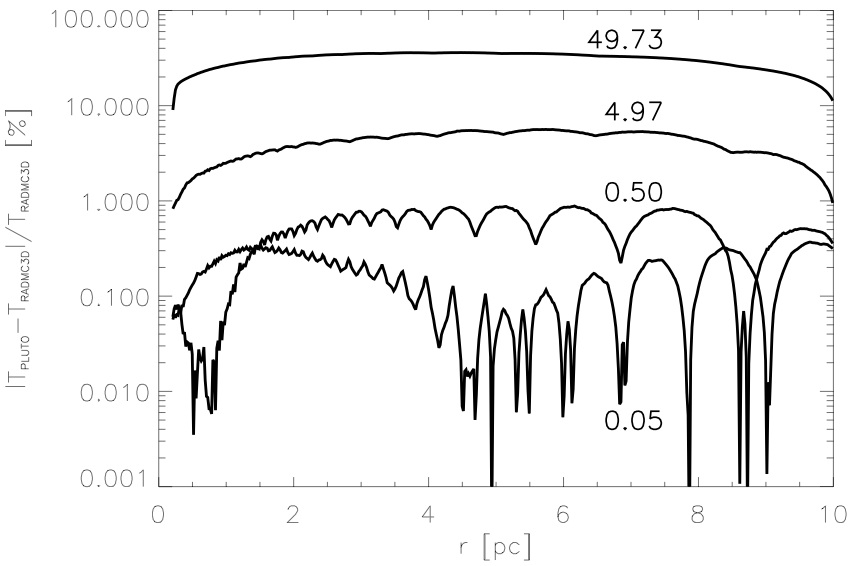}
\caption{Deviation of the solution of our approximate 1D radiative transfer treatment within 
         the PLUTO code \citep{Mignone_07} with
         the resulting temperature distribution derived with RADMC-3D \citep{Dullemond_10}, 
         given in \% for the four optical depths 
         $\tau_{9.8\mu\mathrm{m}}$, as indicated in the plot (compare to Fig.~\ref{fig:plot_comp_radmc3d_pluto}).}
\label{fig:deviation_radmc3d_pluto}
\end{center}
\end{figure}

The resulting radial temperature profiles are plotted in Fig.~\ref{fig:plot_comp_radmc3d_pluto}. 
As expected, our approximate treatment results in lower temperatures
compared to the Monte Carlo approach, as we only take forward reemission into account. 
This effect is stronger for larger optical depths, 
because reemission becomes increasingly important.   
The deviations between the two codes are shown graphically
in Fig.~\ref{fig:deviation_radmc3d_pluto}.
While the optically thin cases mainly display numerical noise, the deviations 
reach the ten percent level in-between the $\tau_{9.8\mu\mathrm{m}}=4.97$ and $49.73$
case. The maxima of the deviations of the radial temperature profiles of the 
test simulations are summarised in Table~\ref{tab:accur_temp}.

\begin{table}
\caption[Accuracy of the temperature distribution]{Accuracy of the radial temperature distribution.}
 \label{tab:accur_temp}
\begin{tabular}{cc}
\hline
$\tau_{9.8\mu\mathrm{m}}$ & $\frac{T-T_{\mathrm{ref}}}{T_{\mathrm{ref}}}$ [\%]\\
\hline
 0.05  & 0.37 \\
 0.50  & 0.88 \\
 4.97  & 5.63 \\
 49.73 & 36.06 \\
\hline
\end{tabular}

\medskip
Maximum deviations of the radial temperature distribution calculated with the one-dimensional approximative
radiative transfer routine implemented in PLUTO \citep{Mignone_07} and the reference calculation, done 
with RADMC-3D \citep{Dullemond_10} for the
four different optical depths tested. 
\end{table}

\subsubsection{Radiation pressure acceleration}

To test the radiation pressure acceleration, we set up a one dimensional dusty shell and illuminate it 
from the centre, without taking gravitational forces into account. 
The domain is set up as in our standard model (see Sect.~\ref{sec:num_setup}). 
The shell is initially located at a radial distance between 4.5\,pc and 5.5\,pc.
We fill the shell homogeneously 
with our standard gas and dust mixture with a density, 
such that the dust optical depths
in radial direction (through the whole model space) lie between 0.1 and 10,000 at a wavelength of 
$9.8\,\mu$m. 

Assuming an optically thick shell of gas and dust and no acceleration mechanism other
than radiation, we analytically 
expect an acceleration of the shell of the following form:

\begin{eqnarray}
\label{equ:radacc}
a_{\mathrm{rad}} = \frac{L_{\mathrm{AGN}}}{4\,\pi\,r^2\,c}\,\frac{1}{N\,\mu}
\end{eqnarray}
where $L_{\mathrm{AGN}}$ is the bolometric luminosity of the central accretion disc, $r$ is the distance to the
centre, $c$ is the speed of light, $N$ the gas column density and $\mu$ the mean particle
mass of the gas. With the assumptions made in this article, the gas column density is directly
related to the optical depth by 

\begin{eqnarray}
\label{equ:ntau}
  N = 9.15\cdot10^{21}\,\mathrm{cm}^{-2}\,\frac{0.60\,m_{\mathrm{a}}}{\mu}\,\frac{16,446.66\,
      \mathrm{cm}^2\mathrm{g}^{-1}}{\kappa}\,\frac{f_{\mathrm{gtd}}}{150.}\,\tau, 
\end{eqnarray}
where $m_{\mathrm{a}}$ is the atomic mass unit and $f_{\mathrm{gtd}}$ is the gas-to-dust mass ratio. 
Given the spherical expansion of the shell, the column density will
scale with $r^{-2}$ and a constant acceleration with time is expected. 

\begin{figure}
\begin{center}
\includegraphics[width=0.9\linewidth]{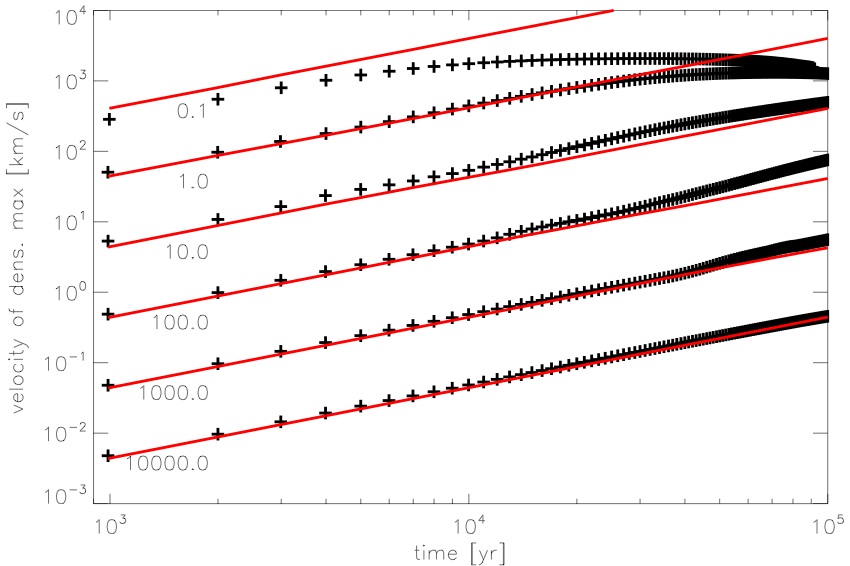}
\caption{Time evolution of the density weighted velocity around the density 
         maximum ($\rho > 0.01\,\rho_{\mathrm{max}}$) of an initially homogeneous shell (black symbols), 
         compared to the analytically expected behaviour (velocity $\propto$ time, thick red lines) for different 
         initial total optical depths (at $\lambda = 9.8\,\mu$m), as annotated.}
\label{fig:shell_velocity}
\end{center}
\end{figure}

We compare this analytical estimate (thick red lines) with the result of our approximative radiation pressure treatment 
(black symbols) in Fig.~\ref{fig:shell_velocity} in terms of the time evolution of the density weighted 
velocity of the shell. For this study, the initial dust optical depth is varied between $\tau_{9.8\mu\mathrm{m}}=$ 0.1 and 10,000, 
as annotated in the figure. For the analytical estimate, the column density is determined from the simulations for each
individual time step.

Very good agreement is found for the highest optical depth case. Towards lower optical depths,
we find two competing effects: (i) in these adiabatic test calculations, pressure effects due to the compression
of the inner boundary layer of the shell increase with decreasing optical depth and get dynamically significant, leading to 
an additional acceleration and
(ii) at some point, the shell gets optically thin and less and less radiation flux is absorbed and can contribute to 
acceleration of the shell. At this point, our analytical estimate is invalid and the goodness of the dynamical evolution
of our approach can be judged by the correctness of the radial temperature distribution (see Fig.~\ref{fig:plot_comp_radmc3d_pluto} 
and \ref{fig:deviation_radmc3d_pluto}). As both of these effects are not present for the case of the highest optical 
depth, the best agreement is found for these cases. 
Thus, we have demonstrated that our approach produces reasonable radial radiative 
accelerations for the simulations presented in this paper. 

\subsection{Numerical Setup}
\label{sec:num_setup}

The initial cloud configuration is chosen to represent a typical clump as seen in the three-dimensional torus simulations
by \citet{Schartmann_10}, as mentioned in Sect.~\ref{sec:introduction}, but with the density adjusted 
to be close to the transition between in- and outflow. Those simulations had been prepared to represent the core region of
the nearby Seyfert~2 galaxy NGC~1068. For the sake of simplicity and to derive the basic behaviour, spherically 
symmetric clouds are assumed, with a homogeneous or initially Gaussian 
density distribution. They are located at a few parsec distance from
the central source and have typically radii of the order of one parsec.
Their kinematics is dominated by radial infall motion and we neglect any overlayed orbital motion 
around the centre. 
The gravitational potential is the same as in \citet{Schartmann_10} and comprises of a supermassive
black hole at the centre with $8\cdot 10^6\,\mathrm{M}_\odot$ and a nuclear star cluster 
with a Plummer potential with a core radius of $25\,$pc and a total mass normalisation constant of 
$2.2\cdot 10^8\,\mathrm{M}_\odot$.
The basic parameters of the 
studies shown in this article are listed and described in Table~\ref{tab:params}.

\begin{table*}
\begin{minipage}{\linewidth}
\caption[Basic parameters of our simulations.]{Basic parameters of our simulations.}
 \label{tab:params}
\begin{tabular}{cccccccccccc}
\hline
name & $r_{\mathrm{initial}}$ [pc] & $R_{\mathrm{cloud}}$ [pc] &
$r_{\mathrm{in}}$ [pc] & $r_{\mathrm{out}}$ [pc] & $\sigma_{\mathrm{c}}$ [pc] 
& $n_{\mathrm{res}}$ & $\rho_{\mathrm{cloud}} [\frac{\mathrm{g}}{\mathrm{cm}^3}]$ 
& $\rho_{\mathrm{min}} [\frac{\mathrm{g}}{\mathrm{cm}^3}]$ 
& EOS & $\epsilon_{\mathrm{edd}}$ & $\tau_{9.8\mu\mathrm{m}}$ \\
\hline
SC00 & 5 & 1 & 0.2  & 10 & 0 & 512 & 1.0e-19 & 1.0e-23 & cool. & 0.1  &  67.4 \\
SC01 & 5 & 1 & 0.2  & 10 & 0 & 512 & {\bf 5.0e-20} & {\bf 5.0e-24} & cool. & 0.1  &  {\bf 33.7} \\
SC02 & 5 & 1 & 0.2  & 10 & 0 & 512 & {\bf 2.5e-20} & {\bf 2.5e-24} & cool. & 0.1  &  {\bf 16.8} \\
SC03 & 5 & 1 & 0.2  & 10 & 0 & 512 & 1.0e-19 & 1.0e-23 & {\bf isoth.} & 0.1 &  67.4 \\
SC04 & 5 & 1 & 0.2  & 10 & 0 & 512 & 1.0e-19 & 1.0e-23 & {\bf adiab.} & 0.1 &  67.4 \\
SC05 & 5 & 1 & 0.2  & 10 & {\bf 0.25} & 512 & 1.0e-19 & 1.0e-23 & cool. & 0.1  &  {\bf 21.2} \\
SC06 & 5 & 1 & 0.2  & 10 & {\bf 0.50} & 512 & 1.0e-19 & 1.0e-23 & cool. & 0.1  &  {\bf 40.5} \\
SC07 & 5 & 1 & 0.2  & 10 & {\bf 1.00} & 512 & 1.0e-19 & 1.0e-23 & cool. & 0.1  &  {\bf 57.7} \\
SC08 & 5 & 1 & 0.2  & 10 & 0 & 512 & 1.0e-19 & 1.0e-23 & cool. & {\bf 0.2}  &  67.4 \\
SC09 & 5 & 1 & 0.2  & 10 & 0 & {\bf 256} & 1.0e-19 & 1.0e-23 & cool. & 0.1  &  {\bf 68.7}\\
SC10 & 5 & 1 & 0.2  & 10 & 0 & {\bf 1024} & 1.0e-19 & 1.0e-23 & cool. & 0.1 &  {\bf 67.7} \\
SC11 & 5 & 1 & 0.2  & {\bf 20} & 0 & 512 & {\bf 2.5e-20} & {\bf 2.5e-24} & cool. & 0.1  &  {\bf 16.9} \\
TC00 & {\bf 5 \& 8} & 1 & 0.2  & 10 & 0 & 512 & {\bf 5.0e-20} & {\bf 5.0e-24} & cool. & 0.1  &  {\bf 33.7 \& 34.0} \\
\hline
\end{tabular}

\medskip
$r_{\mathrm{initial}}$ is the initial distance of the cloud centre from the black hole, $R_{\mathrm{cloud}}$ is the cloud radius,
$r_{\mathrm{in}}$ and $r_{\mathrm{out}}$ are the inner and outer radius of the computational domain,
$\sigma_{\mathrm{c}}$ is the cloud density concentration parameter in case of a Gaussian distribution (zero for a constant density cloud), 
$n_{\mathrm{res}}$ is the number of resolution elements in radial and theta direction, $\rho_{\mathrm{cloud}}$ is the gas density of the cloud, 
$\rho_{\mathrm{min}}$ is the lower gas density threshold in the simulation, EOS is the equation of state and 
$\epsilon_{\mathrm{edd}}$ is the Eddington ratio of the central radiation source, $\tau_{9.8\mu\mathrm{m}}$ is the optical 
depth through the centre of the initial cloud at $9.8\,\mu$m. Bold face indicates parameter changes with respect to our 
standard model.
\end{minipage}
\end{table*}

To evolve the hydrodynamical equations, we use the fully parallel high-resolution 
shock-capturing scheme {\sc PLUTO} \citep{Mignone_07}. For the calculations 
shown in this paper, the two-shock Riemann solver was chosen
together with a linear reconstruction method and directional splitting. 
Optically thin cooling is included with the help of an effective cooling curve for solar metallicity 
(see Fig.~1 in \citet{Schartmann_09} and text therein). All boundary conditions are set to outflow, not
allowing for inflow. We do not take magnetic fields into account in these calculations.
In these two-dimensional simulations, we cannot investigate the dependence of the radiation pressure 
effects on cloud rotation (see discussion in Sect.~\ref{sec:discussion}).

\section{Results}
\label{sec:results}

\subsection{Cloud evolution}
\label{sec:cloud_evolution}

\begin{figure}
\begin{center}
\includegraphics[width=0.8\linewidth]{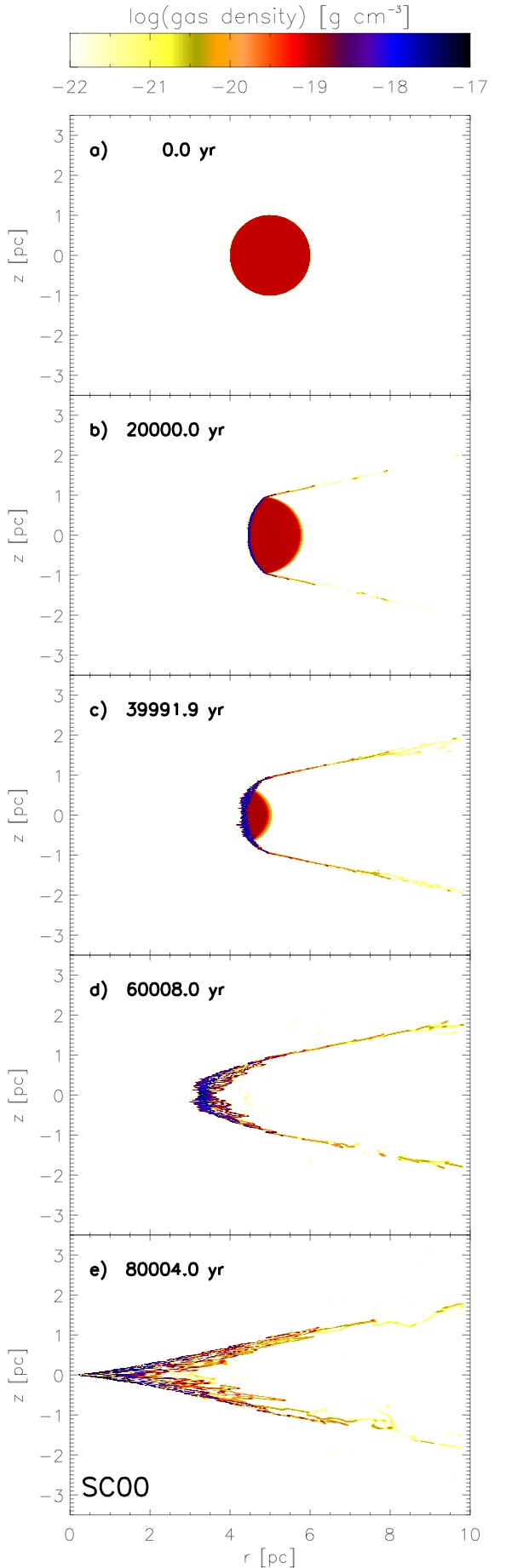}
\caption{Time evolution of the gas density distribution for our standard model (simulation SC00). 
Labels are given
in parsec, the time is given in years. The cloud gets initially compressed and
moves inwards.}
\label{fig:dens_evol_cart_sim000}
\end{center}
\end{figure}

Fig.~\ref{fig:dens_evol_cart_sim000} and 
the upper row of Fig.~\ref{fig:dens_study_evol_dens2d} show the time evolution of the density
distribution for our standard model (SC00) in Cartesian as well as spherical coordinates. 
After having switched on the central radiation source, the initially spherical cloud 
(Fig.~\ref{fig:dens_evol_cart_sim000}a) contracts in radial 
direction (Fig.~\ref{fig:dens_evol_cart_sim000}b). 
The inner boundary of the cloud experiences direct radiation pressure interaction,
which produces a nearly isothermal shock wave with a compression factor
of about a hundred. 
The outer part of the
cloud, which is well shielded from the central radiation source still experiences
the gravitational forces and is accelerated inwards. 
These two effects together result in a converging flow, which causes the formation of
density fluctuations by a number of fluid instabilities (Fig.~\ref{fig:dens_evol_cart_sim000}c). 
Most important in this
case is the non-linear thin shell instability (NLTSI), the
Kelvin-Helmholtz-instability (KHI) and the thermal instability. For a detailed description
we refer to \citet{Heitsch_06}.
As the pressure in the cloud rises above the ambient pressure, the cloud expands in the direction
perpendicular to the radial direction. This gas, together with other low density gas at the 
upper and lower cloud edge is stripped and 
forms long, radial tails which are subject to Kelvin-Helmholtz-instabilities
(Fig.~\ref{fig:dens_evol_cart_sim000}d,e).
These, together with shielding effects lead to the formation of a turbulent wake and some 
mixing of higher density clumps into the shadow region of the cloud. The onset of turbulence is
suppressed in regions with direct lines-of-sight towards the radiation source, as this relatively
low density material suffers from strong outward acclerations. 
With the given parameters, the centre of mass (COM) of the cloud starts moving inwards. 
As radial rays with lower 
column densities suffer from a higher radiation pressure acceleration (Equ.~\ref{equ:radacc}), they lag behind.
This leads to the formation of a narrow, sickle shaped structure (Fig.~\ref{fig:dens_evol_cart_sim000}c,d).
An additional structuring effect results from the gas cooling. 
Dense regions cool on shorter timescale ($t_{\mathrm{cool}}\propto \frac{1}{\rho}$), 
leading to further contraction. As a result of this cooling instability, 
density inhomogeneities within the converging flow are able to contract further, finally forming small
cloudlets of high density material. 
As the acceleration critically depends on the column density of 
the material (see equation\,\ref{equ:radacc}), 
the shell is now able to spread out again (Fig.~\ref{fig:dens_evol_cart_sim000}d,e), 
forming radially extended filaments and high 
density knots, which show a strong dependence on the balance between gravitational and radiation pressure
acceleration forces. In principle, each cloudlet now goes through an evolution similar to our initial cloud. 

In summary, three different phases of cloud evolution are observed: (i) radiation pressure and gravitational
forces lead to a compression of the cloud in radial direction, leading to a lenticular shape (the 
{\it lense phase}, Fig.~\ref{fig:dens_evol_cart_sim000}b,c), (ii) the converging
flows at the inner edge together with cooling of the gas lead to a clumpy, sickle-shaped 
distribution (the {\it clumpy
sickle phase}, Fig.~\ref{fig:dens_evol_cart_sim000}c,d) and (iii) the clumps
induce a {\it column density instability}, which stretches the cloudlets into long radial filaments
(the {\it filamentary phase}, Fig.~\ref{fig:dens_evol_cart_sim000}d,e).

\begin{figure}
\begin{center}
\includegraphics[width=0.9\linewidth]{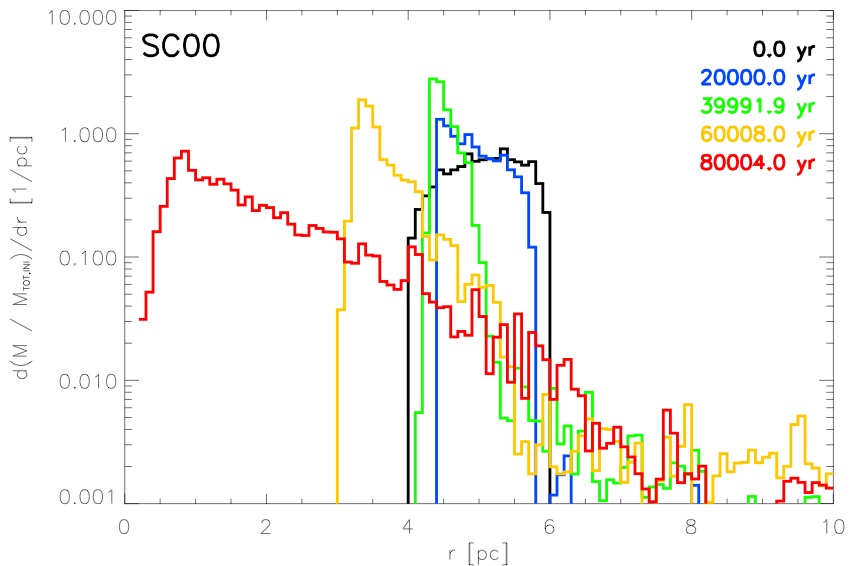}
\caption{The evolution of the distribution of mass onto spherical shells 
is shown for model SC00 for five different time snapshots, as indicated in the legend.}
\label{fig:sc00_massonshells}
\end{center}
\end{figure}

Fig.~\ref{fig:sc00_massonshells} shows the evolution of the distribution of mass onto spherical 
shells (given as a fraction of the initial total mass in the computational domain) in this model. The initial 
distribution is given in black, whereas the other colors are used 
for later snapshots, as indicated in the legend of the plot. It quantifies the behaviour as already 
described above: first of all, radiation pressure leads to a density maximum at the inner boundary, 
whereas the unaffected cloud still gets accelerated towards the centre, leading to a converging flow.  
Altogether, the mass distribution gets narrower in radial direction (concerning the bulk of the gas). 
As soon as the clumpy structure 
has evolved, the various column densities in radial direction lead to slightly varying radial velocities
for different theta angles. This can be seen as an increase of the width of the profile, as visible
for the case of the red curve. By that time (80,000\,yr after the start of the simulation), 
parts of the filaments have already reached the sublimation radius, which approximately 
coincides with the inner boundary in our simulations. 

\begin{figure}
\begin{center}
\includegraphics[width=0.9\linewidth]{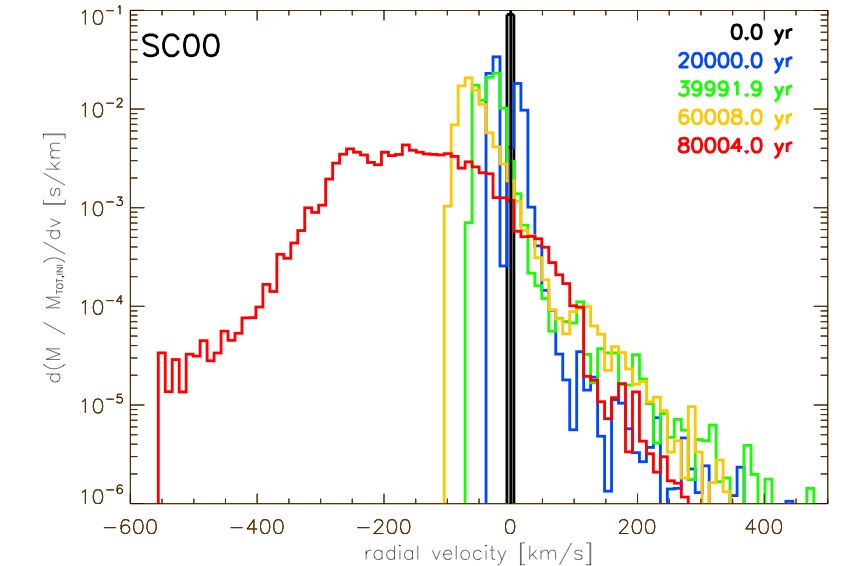}
\caption{Mass histogram of the radial velocity distribution
         of our standard model (SC00).}
\label{fig:sc00_radvel_hist}
\end{center}
\end{figure}

\begin{figure*}
\begin{center}
\includegraphics[width=0.9\linewidth]{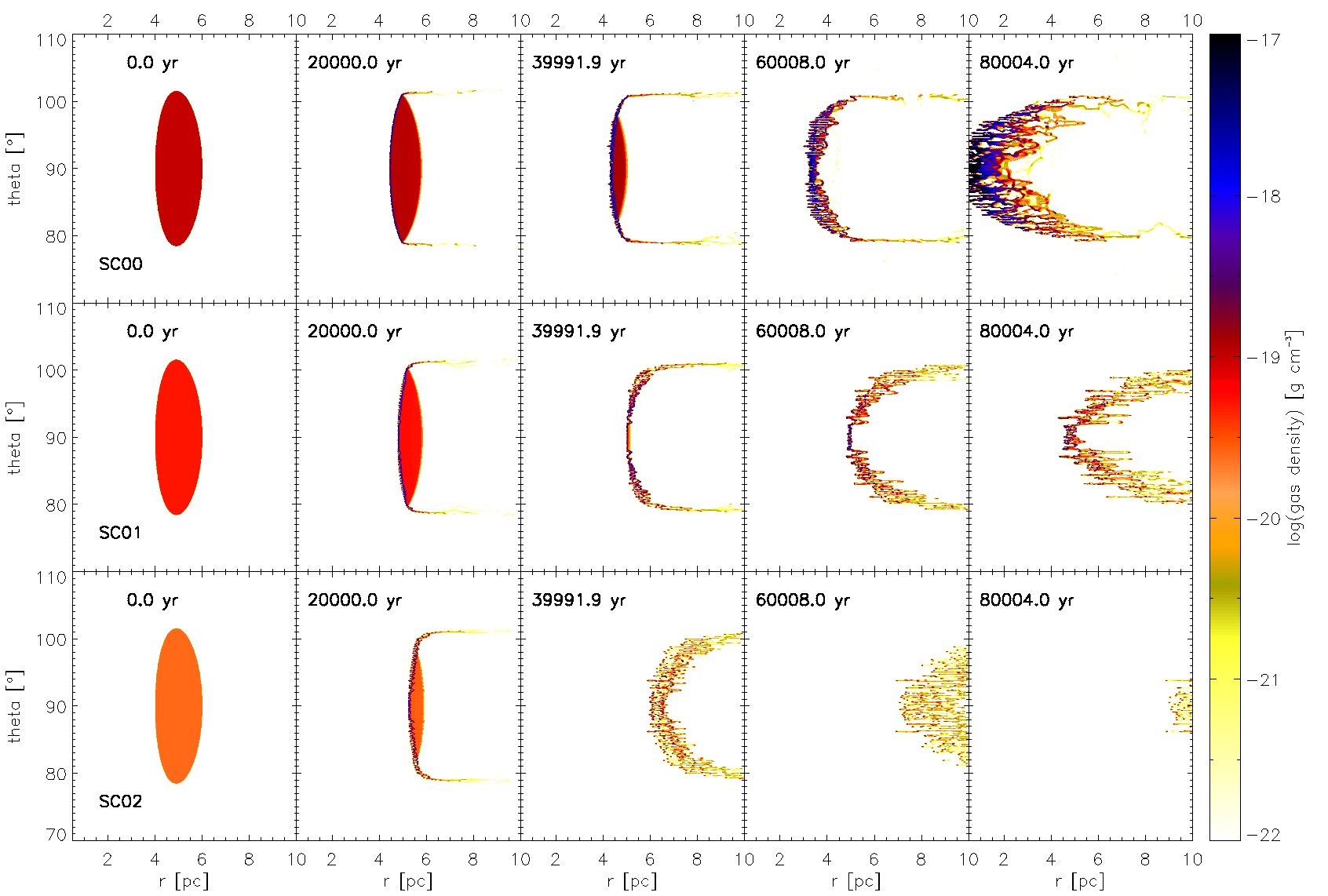}
\caption{Density evolution for clouds with various masses. Shown are the three 
         simulations SC00 ($\rho_{\mathrm{cloud}}=1\cdot 10^{-19}$g\,$\mathrm{cm}^{-3}$, upper row), 
         SC01 ($\rho_{\mathrm{cloud}}=5\cdot 10^{-20}$g\,$\mathrm{cm}^{-3}$, middle row) and 
         SC02 ($\rho_{\mathrm{cloud}}=2.5\cdot 10^{-20}$g\,$\mathrm{cm}^{-3}$, lower row).}  
\label{fig:dens_study_evol_dens2d}
\end{center}
\end{figure*}

The differential gas mass distribution as a function of the radial velocity is shown in 
Fig.~\ref{fig:sc00_radvel_hist}. Initially, the cloud is at rest (black line). The various colours 
show different timesteps as indicated in the legend. With time, the distribution broadens. The bulk of 
the mass is accelerated inwards, whereas a small fraction (at the low column density edges of the cloud) 
gets accelerated outwards. This is again caused by the vertical small scale variations of the column density.
Most of the mass is contained in the inward accelerated dense cloudlets, 
whereas the low density tails contribute only little to the total 
mass budget.The cutoff at large infall velocities is caused by the fact, that part of the material has 
already left the model space.

\subsection{Cloud density study}

Fig.~\ref{fig:dens_study_evol_dens2d} shows the time evolution of the density distribution of a cloud density study, 
displayed in spherical coordinates. 
The cloud's column density was chosen such that the clouds are close to force equilibrium between
the gravitational and radiation pressure force.
The first row corresponds to the standard model (SC00) discussed in 
Sect.~\ref{sec:cloud_evolution}. For the second and third row, we halved (SC01) and quartered (SC02) the density 
of the cloud respectively. 
Within the density range shown, all three basic evolutionary phases discussed above can be recognized, 
but several distinct differences exist: 
As expected, the motion of the centre of mass changes and the inward motion is slower for lower density values and reaches a 
transition to outflow after a column density threshold is reached (see Sect.~\ref{sec:ana_exp}).
For the case of the outflowing cloud (SC02), we also see a larger spread of the sickle-shaped cloud in radial direction.
For the case of even higher column densities than displayed, the amount of compression decreases, as gravitational acceleration 
dominates over radiation pressure effects. Radiation pressure effects are only
effective at the cloud edges, where cometary-shaped tails build up, which develop Kelvin-Helmholtz-instabilities
and form a turbulent wake behind the cloud. Only the inner boundary of the cloud forms small cloudlets and filaments, whereas
the outer part keeps a continuous density distribution. On its way towards the centre, the cloud shrinks due to 
gravitational forces, gas cooling and the ablation of gas at the outer edges.

In summary, clouds can encounter two fates, depending on their radial column density:  
(i) Low column density clouds will be pushed outwards, whereas 
(ii) high column density clouds always show both, in- and outflow motion, due 
to the formation of tails at the low column density edges of the clouds.

\begin{figure}
\begin{center}
\includegraphics[width=0.9\linewidth]{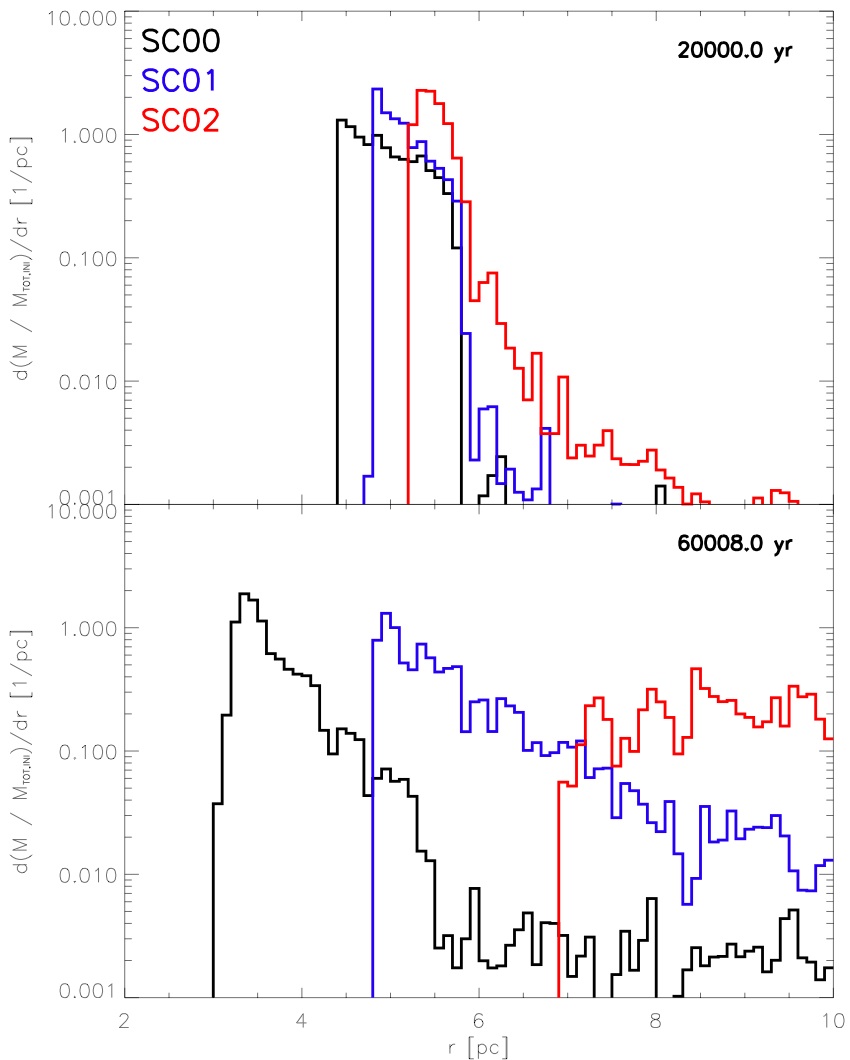}
\caption{Distribution of mass on spherical shells for a density study. Our standard model (SC00)
is given in black, whereas the blue line corresponds to an initial cloud mass of half the standard
model's value (SC01) and the red line to a quarter of it (SC02). 
It is shown for two time snapshots in the upper and lower panel
as indicated in the upper right corner.}
\label{fig:dens_study_massonshells}
\end{center}
\end{figure}

Fig.~\ref{fig:dens_study_massonshells} shows the radial distribution of mass for clouds with 
different densities for an early 
snapshot (upper panel, after 20,000\,yr) and a late snapshot (lower panel, after 60,000\,yr). 
It clearly shows the dependence of the density distribution on the initial condition: 
whereas the high density case remains peaked, it spreads out more and more for the lower column density clouds.

\begin{figure}
\begin{center}
\includegraphics[width=0.9\linewidth]{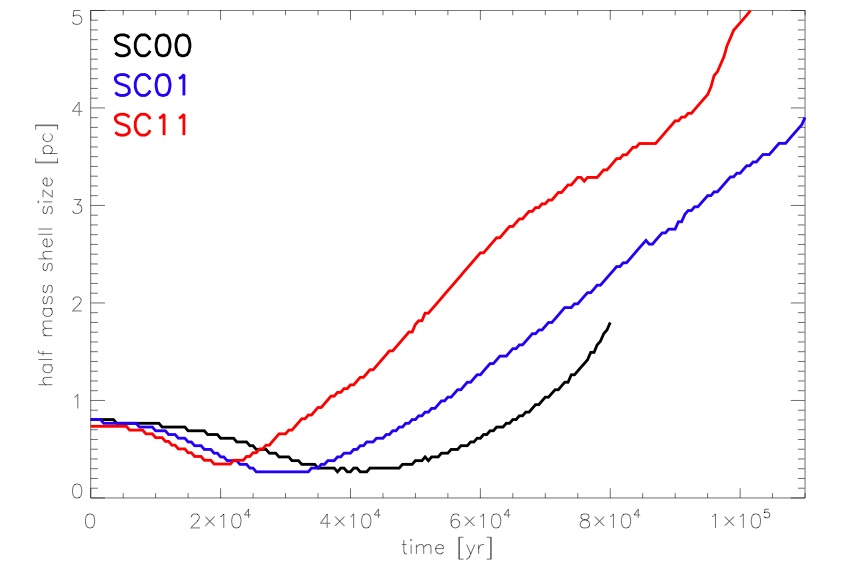}
\caption{Time evolution of the thickness of the concentrical shell around the 
origin at the location of the centre of mass, which encloses 50\% of the 
initial mass. SC00 is our standard model, SC01 has half the mass of the 
standard model and SC11 a quarter of the mass. SC11 is identical to SC02, but
we increased the outer radius of the computational domain to 20\,pc.}
\label{fig:density_study_shell_thickness}
\end{center}
\end{figure}

\begin{figure}
\begin{center}
\includegraphics[width=0.9\linewidth]{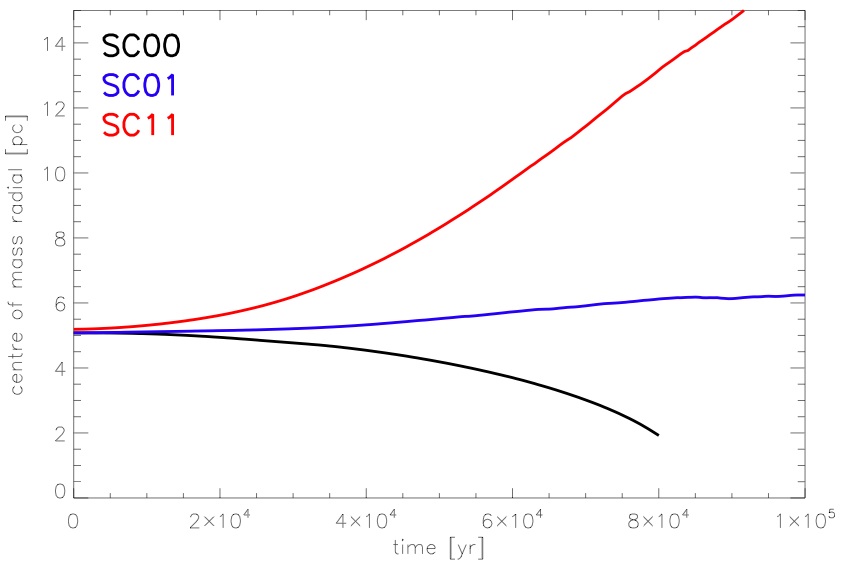}
\caption{Time evolution of the radial location of the centre of mass for the 
cloud density study. SC00 is our standard model, SC01 has half the mass of the 
standard model and SC11 a quarter of the mass. SC11 is identical to SC02, but
we increased the outer radius of the computational domain to 20\,pc.}
\label{fig:density_study_cloud_com_r}
\end{center}
\end{figure}

The shearing of the cloud is quantified in Fig.~\ref{fig:density_study_shell_thickness}.
First of all, we determine the centre of mass of the cloud. Then we add up the mass in the spherical shell defined 
by the radial cell, which contains the COM. We extend the spherical shell (symmetrically with respect to
the centre of mass) until it includes 50\% of the mass of the initial condition within the spherical shell embracing
the initial cloud.  
This is done for the standard model (SC00, black line), the model with half the mass of the 
standard model (SC01, blue line) and a model with a quarter of the mass of the standard model with an
outer radius extended to 20\,pc (SC11, red line). 
In the beginning of the simulation, the contraction of the cloud is visible, before the differential
forces lead to a shearing. Two types of shearing occur: (i) due to the column density differences between the
cloud centre and the cloud's outer edge, as visible in the extended tails 
and (ii) due to the small scale column density differences 
which emerge in a later stage of the evolution.
After the compression phase, the half mass shell size increases almost linearly for the case where the
COM moves outward (see Fig.~\ref{fig:density_study_cloud_com_r}). 
The evolution is fastest for the low density case, where radiation pressure
dominates the radial forces. The infalling high density case
behaves differently, as the dense inner shell which forms due to the initial contraction seems to 
prevent efficient shearing.  

\begin{figure}
\begin{center}
\includegraphics[width=0.9\linewidth]{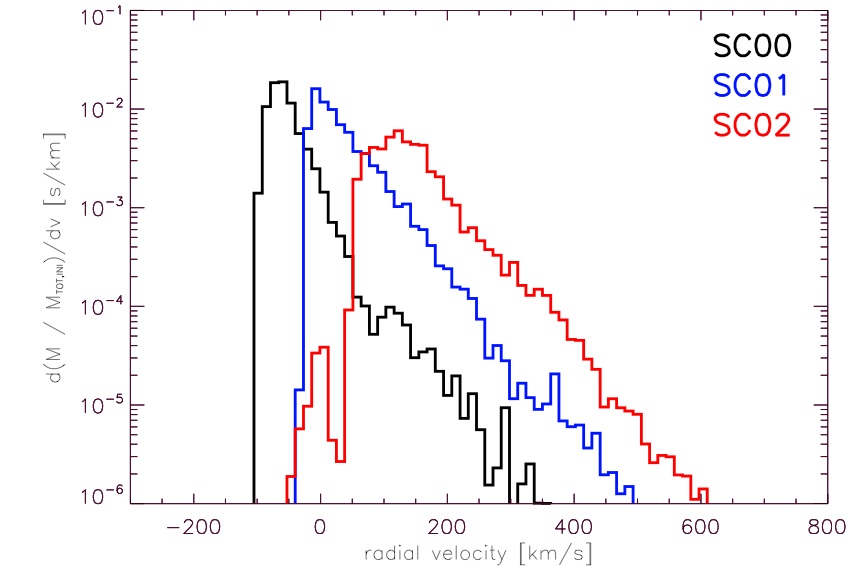}
\caption{Histograms of the distribution of mass into bins of radial velocity for the cloud density study, after an
evolution time of 60,000\,yr (SC00 -- standard model, SC01 -- half the mass, 
SC02 -- a quarter of the mass).}
\label{fig:dens_study_radvel_hist}
\end{center}
\end{figure}

Fig.~\ref{fig:dens_study_radvel_hist} quantifies the different velocities reached. 
Whereas the highest density cores reach the smallest outflow and the highest
inflow velocities, respectively, the highest outflow velocities are reached by the material within the tails expelled from the 
cloud edges. This material shows up as the extended tails of the distributions in Fig.~\ref{fig:dens_study_radvel_hist}.

\begin{figure*}
\begin{center}
\includegraphics[width=0.8\linewidth]{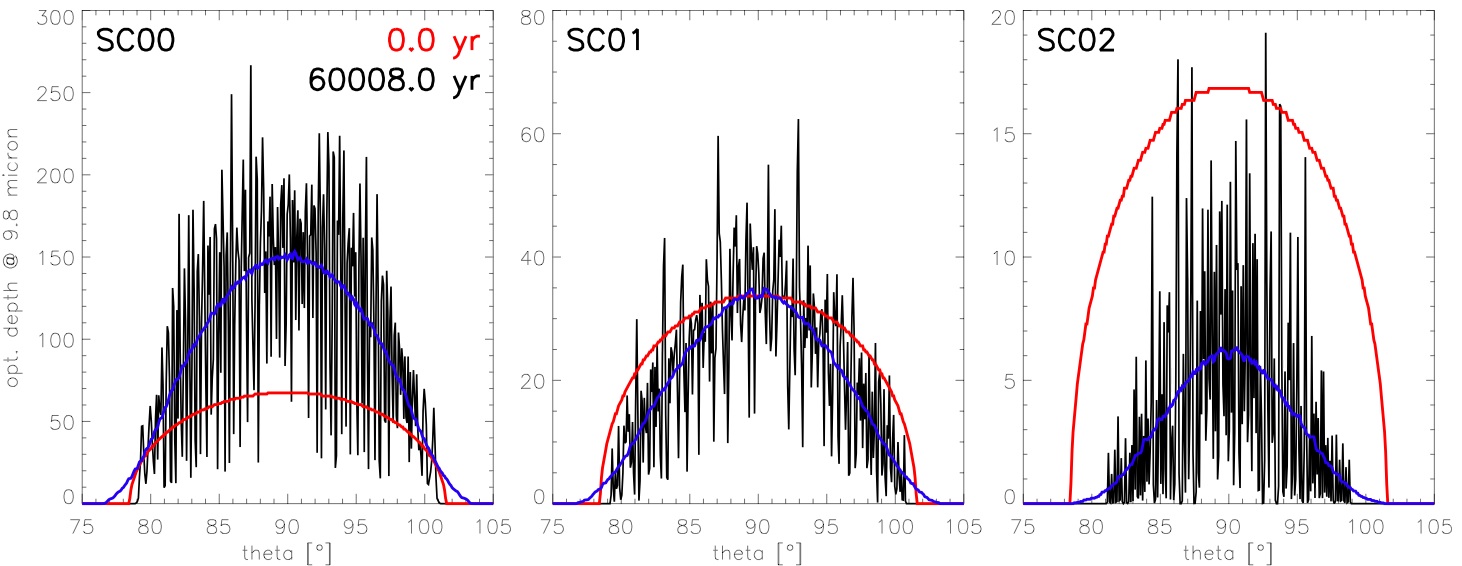}
\caption{Distribution of the optical depth in radial direction. Shown are the initial distribution (red lines) and
the distribution after roughly 60,000\,yr (black lines) for the density study 
(SC00 -- standard model, SC01 -- half the mass, SC02 -- a quarter of the mass). The blue line represents the distribution
after 60,000\,yr, but smoothed with a boxcar function of 5 degrees width in theta direction.}
\label{fig:dens_study_opdep_evol}
\end{center}
\end{figure*}

Fig.\,\ref{fig:dens_study_opdep_evol} shows the evolution of the optical depth (or the column density distribution, 
compare to Equ.\,\ref{equ:ntau}) in radial direction for the three 
simulations of the gas density study. Each panel displays the initial distribution in red and the state after 
roughly 60,000\,yr in black. The overlayed blue line corresponds to the black line, but is smoothed with a boxcar 
function of 5 degrees width in theta direction. In all three cases, the distribution gets more peaked with respect  
to the initial distribution. The high density case shows a global increase of the optical depth, while it stays 
almost constant in simulation SC01 and overall decreases for the case of simulation SC02. Both of these effects are
mostly caused by the inward or outward motion of the cloud. Being accelerated inward leads to a contraction of the cloud 
by gravitational forces, whereas the outward moving cloud expands due to the radially acting radiation pressure
forces. As already discussed, the cloud evolves into a sickle shape. Being stretched radially, the edges rather 
expand, whereas the central parts rather contract, leading to a more peaked distribution of the optical depth. 
The standard deviation of the distribution of the later snapshot with respect to the smoothed curve drops from
24\% of the peak value of the smoothed curve for the high density case to 19\% for the intermediate density case and 
amounts to 44\% for the low density case. 
The intermediate density case is closest to the equilibrium between gravitational and radiation pressure forces 
leading to only minor inward and outward motion in the early phase. In the other cases, contraction or expansion of the whole 
cloud leads to enhancement of the column density fluctuations.

\subsection{Comparison to analytical expectations}
\label{sec:ana_exp}

In this section, we compare the general dynamics of the cloud remnant to 
a simple analytical acceleration model. 
For the latter, we assume that the clouds retain their initial cross section  
throughout the whole acceleration process. The cloud 
absorbs all incident radiation onto its geometrical cross section of 
$\sigma=\pi\,R_{\mathrm{cloud}}^2$. Including the gravitational attraction 
of the central black hole and the stellar cluster, this results in an
accelerating force of
\begin{eqnarray}
  F_{\mathrm{acc}} = \frac{L_{\mathrm{AGN}}}{4\,\pi\,r^2\,c}\,\sigma - \frac{G\,m_{\mathrm{cloud}}\,M(r)}{r^2},
\end{eqnarray}

where $L_{\mathrm{AGN}}$ is the luminosity of the central accretion disc, $M(r) = M_{\mathrm{BH}} + 
\hat{M}_*\,\frac{\left(\frac{r}{r_{\mathrm{c}}}\right)^3}
{\left(1+\left( \frac{r}{r_{\mathrm{c}}} \right)^2 \right)^{3/2}}$ is 
the mass of the central supermassive black hole and the nuclear star cluster. The latter
is modelled as a Plummer profile with core radius $r_{\mathrm{c}}$. 
$G$ is the gravitational constant and $c$ the vacuum speed of light. A second 
order accurate leapfrog algorithm is used to integrate the dynamics of the cloud. 

\begin{figure}
\begin{center}
\includegraphics[width=0.9\linewidth]{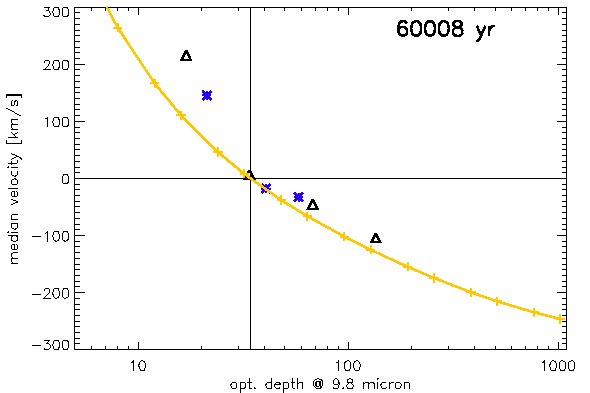}
\caption{Comparison of our RHD simulations (black triangles for the density study, blue stars for the 
         Gaussian distribution study, see Table~\ref{tab:params}) and a simple acceleration model 
         of spherical clouds (yellow line and symbols).}
\label{fig:inoutflow_transition}
\end{center}
\end{figure}

In Fig.~\ref{fig:inoutflow_transition} we compare the result of this experiment with our 
simulations. The velocity of the cloud or cloud remnant is shown after an evolution 
time of roughly 60,000 years for various values of the optical depth through the midplane of 
the cloud. For the case of the numerical radiation hydrodynamics (RHD) simulations, both values are determined by 
averaging over 10 grid cells in vertical direction, centered on the midplane. 
In radial direction, all grid cells, which are filled with dust are taken into account when summing up
the optical depth and calculating the median velocity, respectively.
Plotted in yellow is the result of the analytical estimate. The black triangles and blue stars refer to our
RHD simulations listed in Table~\ref{tab:params}.

In general, the simulations are in reasonable agreement with the simple analytical estimate. As expected,
the agreement is better for the high optical depth case. 
For decreasing optical depth values, smaller and smaller areas of the clouds fulfill
the assumption that all lines of sight through the clouds are optically thick.
The deviations arise from the fact that in the real hydrodynamical simulations, the clouds get disrupted 
by the action of the radiation pressure and gravitational forces together with the onset of instabilities,
leading to overall larger sizes and, therefore, smaller column densities and higher outflow velocities.

\subsection{Parameter studies}

In the following we summarise the main results from several parameter studies:

\begin{figure*}
\begin{center}
\includegraphics[width=0.9\linewidth]{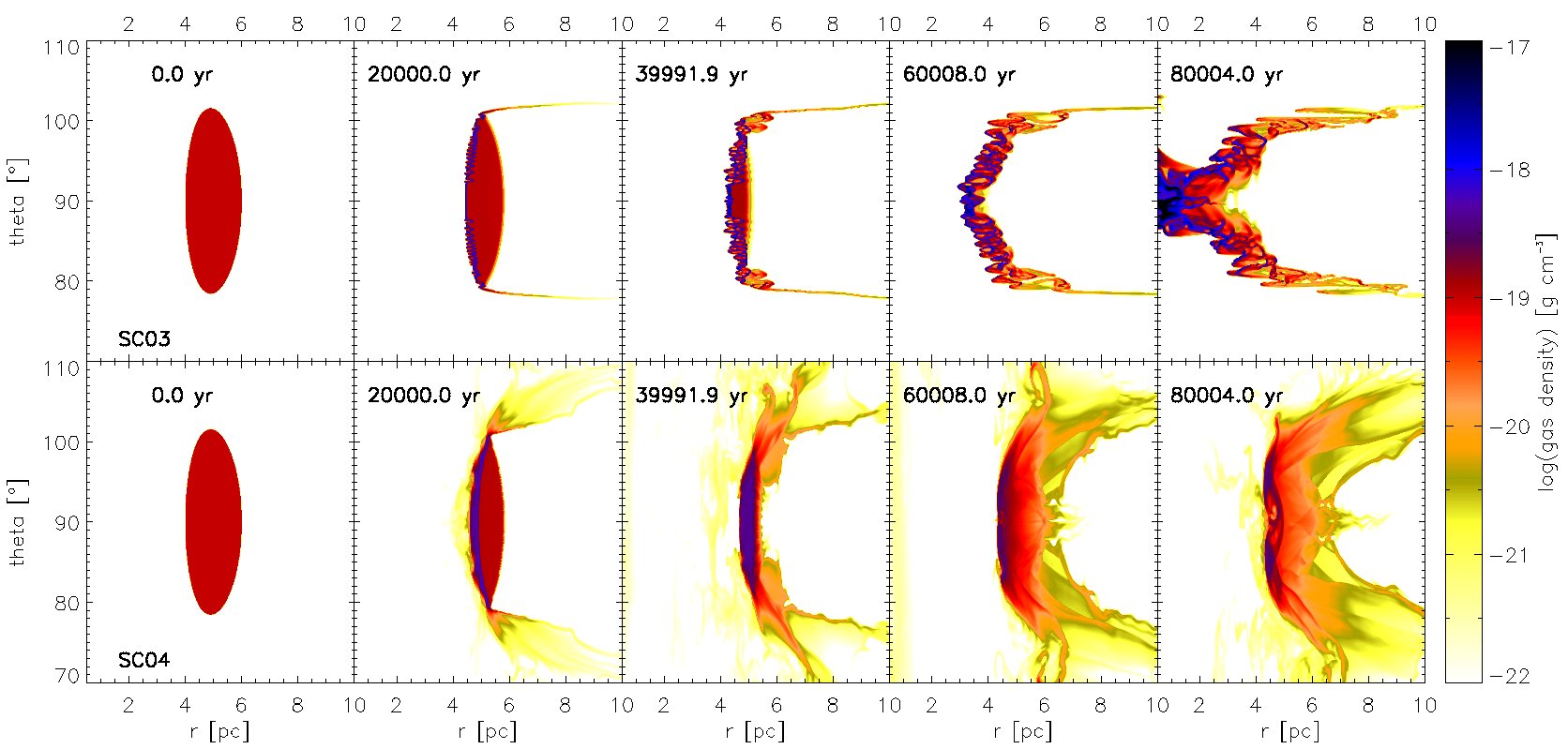}
\caption{Comparison of the evolution of the density for an isothermal simulation (upper row, SC03)
         and an adiabatic simulation (lower row, SC04). The parameters are as in our standard 
         simulation (SC00, see Fig.~\ref{fig:dens_evol_cart_sim000} and upper row of 
Fig.~\ref{fig:dens_study_evol_dens2d}).}
\label{fig:eos_study_dens2d}
\end{center}
\end{figure*}

\begin{enumerate}
  \item {\bf Equation of state:}

For this study, we use the same setup as in our standard model but adopt
an isothermal equation of state for each of the two temperature phases of 
our initial condition or an adiabatic 
equation of state. The results show substantial differences (see
Fig.~\ref{fig:eos_study_dens2d}): In the isothermal simulation (upper row, SC03), 
the converging flow produces a filamentary or clumpy structure as well, but with 
larger characteristic sizes due to the higher remaining thermal pressure in the cloudlets. 
This leads to smaller column densities and the cloud reaches smaller 
infall velocities compared to the case where we take gas cooling into account
(Fig.~\ref{fig:dens_evol_cart_sim000}). 
In the 
adiabatic simulation, which resembles the situation of inefficient gas cooling
and efficient gas heating processes, the gas is heated to temperatures above the dust sputtering 
threshold at the inner edge of the cloud and dust is destroyed. Parts of the cloud are able to evaporate and
leave the cloud, as they are not subject to radiation pressure forces anymore in our
scheme. Due to the compression in the rim of the cloud, it is now overpressured with
respect to the surrounding medium, leading to a slight expansion in vertical direction.
As a consequence of the high pressure, the density distribution remains much 
smoother compared to the cooling case and no strong clumping can occur.
Given the lower column density, the cloud's centre of mass is slightly pushed outwards
during the runtime.

\item {\bf Gaussian internal structure:}

For this study, we set up a 
Gaussian internal density distribution 
($\rho = \rho_0 \, \exp{-\frac{(x-r_{\mathrm{initial}})^2}{2\,\sigma_{\mathrm{c}}^2}}$), 
where $\rho_0$ is the value of the gas density in the centre of the cloud and
we vary the concentration parameter $\sigma_{\mathrm{c}}$ from $0.25$\,pc to $0.5$\,pc 
and $1$\,pc. The stronger the concentration of the mass of the cloud towards the
centre, the stronger is the compression of the sickle shape, as the low density
outer regions can be pushed outward easily.
The cloud centre of mass behaves 
like expected from the initial midplane column density, which decreases with stronger 
concentration according to our definition (see Table\,\ref{tab:params}). This can be seen in 
Fig.~\ref{fig:inoutflow_transition}. The Gaussian
internal structure study is shown here as the blue stars.

\item {\bf Eddington Ratio:}

\begin{figure}
\begin{center}
\includegraphics[width=0.8\linewidth]{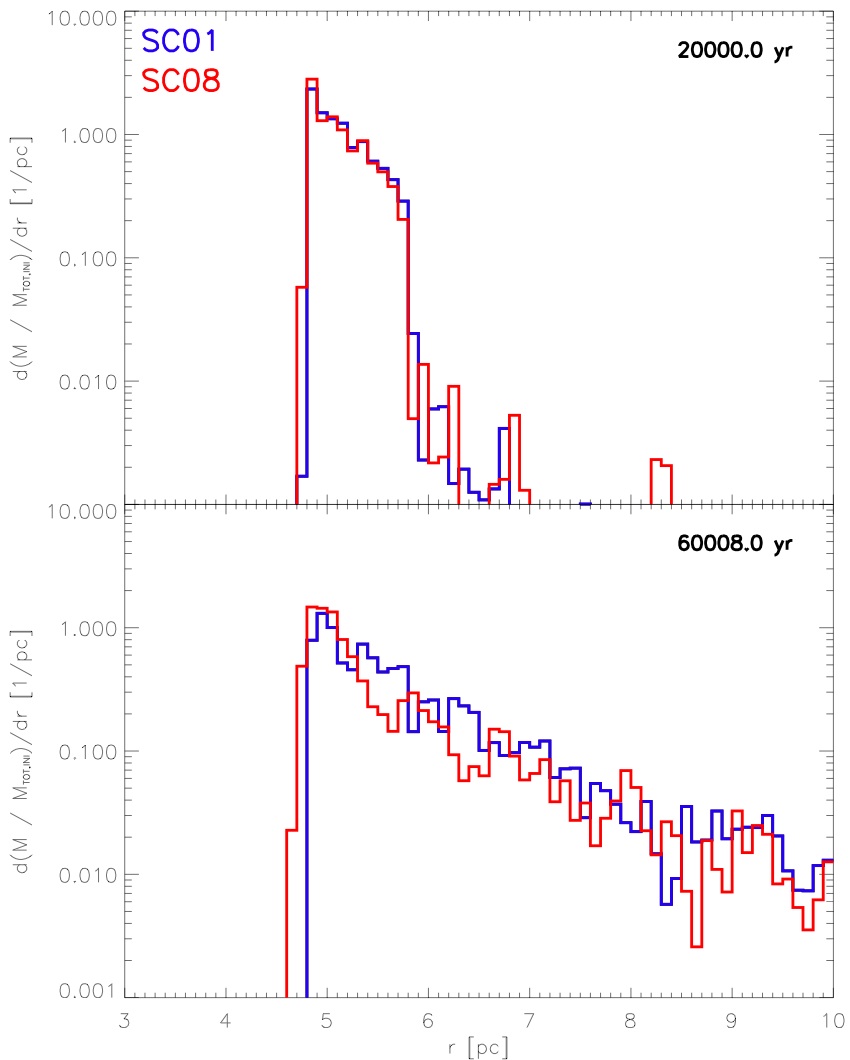}
\caption{The distribution of mass onto spherical shells is compared between model SC01 (blue graphs), where the 
cloud density is half the density of the standard model and SC08 (red graphs), where the luminosity of the 
central source is twice the luminosity of our standard model. The two panels correspond to two different timesteps.}
\label{fig:dens_eddratio_massonshells}
\end{center}
\end{figure}

As can be seen from equation \ref{equ:radacc}, the radiation pressure acceleration in the optically 
thick limit scales proportional to the source luminosity $L_{\mathrm{AGN}}$ and to the inverse of the
gas column density. Therefore, for the case of the Eddington ratio study, we expect a similar
behaviour as for the density study. This is indeed the case, 
as can be seen from Fig.~\ref{fig:dens_eddratio_massonshells},
where the simulation with half of the mass of the standard model (SC01, blue line) is compared to
the standard model illuminated with twice the source luminosity (SC08, red line).

\end{enumerate}

\subsection{Interacting clouds}

\begin{figure*}
\begin{center}
\includegraphics[width=0.9\linewidth]{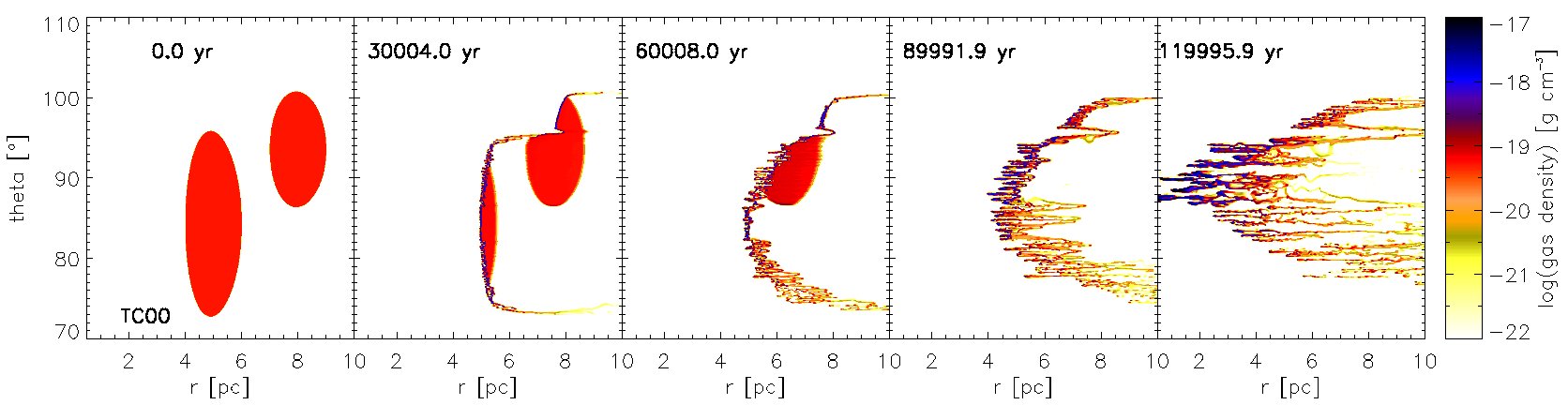}
\caption{Time evolution of the density distribution for the scenario of two interacting clouds (simulation TC00).}
\label{fig:tc01_dens_evol_sph}
\end{center}
\end{figure*}

In order to study two interacting clouds, we use the same cloud characteristics as in model SC01, but
offset the cloud slightly in polar direction. The second cloud (identical to the first one)
is placed at a location of 8\,pc from the centre and offset by one cloud radius in polar direction. 
In Fig.~\ref{fig:tc01_dens_evol_sph}, the evolution of the density for this setup 
is shown for a number of snapshots. 
During the first steps of the evolution, the inner cloud evolves as described in 
Sect.~\ref{sec:cloud_evolution}. 
Being optically thick, it casts a shadow on parts of the 
outer cloud,  which -- in consequence -- reacts only to
the gravitational acceleration in this part. In the transition 
region between shadowed and unshadowed part, the tail of the inner cloud
interacts with the outer cloud, punching a hole into it due to the additional ram pressure. 
For the cloud parameters shown here, the two clumps will collide, leading to
the formation of cloudlets with high enough column density to lead to inflow 
motion.

\subsection{Resolution study and numeric parameters}
\label{sec:numerics}

A resolution study has been conducted where we 
decreased the number of grid cells to $256^2$ (SC09) and 
doubled it to $1024^2$ (SC10) with respect to our standard
model. As expected, the clumpiness of 
the dense shell depends on the resolution. However, the overall behaviour 
of the cloud, concerning acceleration, mass distribution and 
radial velocity is very similar. The higher resolved clouds show a 
slightly broader distribution of mass along the force direction. 
There, the column density instability can act on shorter time scales, 
spreading out the remaining cloud in radial direction. We also 
find a resolution dependence in the maximum density, as expected.

Several other studies have been undertaken in order to investigate the
influence of the numerics on the results of this work. 
Changing the minimum gas temperature of the simulation changes the 
minimum size of the cloudlets. The higher we choose this threshold value,
the smoother the cloud appears. Clumps continue to form due to the
converging flow.

By increasing the density of the surrounding medium, at some point, its 
cooling time is small enough to form small clumps, which overcome the
density threshold for radiation pressure interaction in our simulations. 
With their ram pressure, they lead to a faster formation of filaments in 
the cloud.

\section{Discussion}
\label{sec:discussion}

\begin{figure}
\begin{center}
\includegraphics[width=0.9\linewidth]{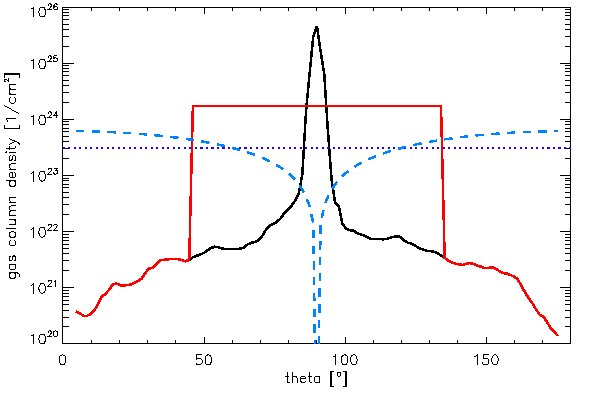}
\caption{Gas column density distribution of the 3D torus simulations presented in \citet{Schartmann_10} (black line), 
         compared to the approximate column density threshold as derived numerically in this article (dark blue dotted line) and taking 
         the $|\cos(\vartheta)|$-radiation characteristic into account (light blue dashed line). For the case of 
         the red line, we distributed the mass of the central disc within a geometrically thick disc structure, which
         brings the distribution well above the threshold.} 
\label{fig:coldens_3d}
\end{center}
\end{figure}

In this paper, we investigated the radiation pressure interaction of infalling dusty gas clouds with
the active nucleus of a Seyfert galaxy, exemplified for the physical parameters of NGC\,1068. 
Dictated by the gas column density, clouds will be accreted or expelled from the central region. 
Outward accelerated clouds will interact and merge with clouds and filaments further out, until
the critical column density is reached.  
Fig.~\ref{fig:coldens_3d} shows the column density distribution in radial direction 
for the 3D model of the Seyfert\,2 galaxy
NGC\,1068 as discussed in \citet{Schartmann_10}. It is shown for all polar angles, each of them
averaged over the azimuthal angle. Clearly visible is the two-component structure -- the geometrically
thin, but high-column density disc and the extended low-column density torus on tens of parsec scale. 
The dark blue dotted line denotes the transition between in- and 
outflow motion as derived in this article. The latter is only a rough estimate, as it neglects 
the radius dependence given by the extended potential of the nuclear star cluster.
The light blue dashed line shows the same column density threshold, but assuming
a radiation characteristic of the source proportional to $|cos(\vartheta)|$. 
According to this column density distribution, most of the 
large scale filamentary torus component would be expelled from the central region, as it is unshielded by 
the central parsec scale disc component. 
The red line refers to a model, where we distributed the material in the disc (up to a distance from the 
black hole of 2\,pc and an angle of $\pm45\degr$ with respect to the disc midplane) in a wedge-shaped structure with
a 45 degree half opening angle. This should resemble the geometrically thick dusty torus, which would result from our
inner disc component, if the scaleheight would have been increased by a thus far unknown turbulent process. 
As can be seen from Fig.~\ref{fig:coldens_3d}, a large fraction of the model space is then sufficiently 
shielded to allow for further feeding of the central galactic region, 
whereas gas far away from the midplane will still be pushed outward.

For all of the simulations shown in this paper, we assume that gas and dust are thermally decoupled. 
This is strictly true only up to a threshold density of $10^{-18}\,\mathrm{g}\,\mathrm{cm}^{-3}$ 
\citep{Larson_05}. At high densities, the gas temperature is given by the 
heating and cooling processes of the dust. This limit is taken into account only very 
roughly in our simulations by setting up a minimum gas temperature, which 
affects the minimal extension of the cloudlets.

In this first set of models, we neglect magnetic fields mainly for the sake of simplicity. 
Despite this, magnetic fields are supposed to be important in these galactic nuclear regions, but their 
strengths and morphologies are basically unknown. In the context of such simulations of dusty clouds, two main effects are expected: 
(i) magnetic fields provide an additional pressure component and (ii) the interaction of magnetic fields with charged ions and dust grains
leads to a strong coupling of the gas and dust phase (e.~g.~\citealp{Scoville_95}). The latter enables us to treat 
the two phases in a one-fluid approximation. The clouds we simulate in this paper contain the products of stellar evolutionary
processes. Many expelled outer atmospheres of AGB stars have merged to form a larger entity. As these stellar atmospheres were 
observationally found to be significantly magnetised (e.~g.~\citealp{Gomez_09} and references therein) by 
stellar dynamos at the interface of the rapidly rotating core and the more slowly rotating envelope \citep{Blackman_01}, 
the clouds will be magnetised as well. 
Depending on the (unknown) morphology of the magnetic field lines, their presence may or may not have a stabilising 
effect on the cloud, which is exposed to radially compressing gravitational and radiative pressure forces. 
The hot and highly ionised ambient medium is also likely to be magnetised, which might provide an additional confining
magnetic pressure, resulting in enhanced stability of the cloud. This scenario was proposed by 
\citet{Rees_87} to be the dominant confinement mechanism for BLR clouds. \\

We also neglect the possibility that clouds are rotating. Depending on the rotation frequency compared to the 
cloud distortion time due to radiation pressure and gravity, this can have a significant effect on the
evolution of the cloud and might lead to a different dynamical behaviour of the centre of mass of the cloud.
However, to study this, three-dimensional simulations are necessary and we defer this to our future work.

\section{Conclusions}
\label{sec:conclusions}

With the help of three-dimensional hydrodynamical simulations considering the effects of stellar evolutionary 
processes of a nuclear star cluster on the surrounding interstellar medium (ISM), \citet{Schartmann_09,Schartmann_10} 
typically find a two-component structure: a geometrically thin, but optically thick 
disc on sub-pc to pc scale, surrounded by a filamentary or clumpy distribution on tens of parsec scale. 
Building up on these results, we investigate the further evolution of these clumps and filaments 
after the activation of the central nucleus, employing a simplified treatment: spherical clouds are 
studied taking gravitational forces, radiative transfer effects and optically thin line cooling
into account.

The evolution of the clouds can be separated into three different phases:
\begin{enumerate}
  \item the {\it lense phase}: counteracting forces (radiation pressure and gravity) transform the cloud into a lenticular shape
  \item the {\it clumpy sickle phase}: due to converging flows and cooling instability, a number of cloudlets form in a sickle-shape,
        whose general structure is dictated by the initial internal column density distribution of the cloud
  \item the {\it filamentary phase}: the strung cloudlets introduce a column density instability, leading to the formation of 
        long radial filaments.
\end{enumerate}

The fate of the clouds ultimately depends on the column density of the matter. 
In summary, two scenarios are possible: (i) low column density clouds will be completely pushed outwards and (ii) high 
column density clouds loose part of their material at the edges, whereas the bulk of the matter moves inwards. The 
general dynamical evolution of the cloud can approximately be described with the help of a simple analytical model, where
the gas column density determines whether the cloud will move inward or will be pushed outward.

\section*{Acknowledgments}
Part of the numerical simulations have been 
carried out on the SGI Altix 4700 HLRB II of the Leibniz 
Computing Centre in Munich (Germany). 
We thank the anonymous referee for his suggestions to 
improve the paper and K.~Tristram and L.~Burtscher for proofreading 
the manuscript.

\bibliographystyle{mn2e}
\bibliography{astrings,literature}

\end{document}